\documentclass[sigconf]{acmart}

\newtheorem{theorem}{Hypothesis}
\newtheorem{theoremgrp}{Hypothesis}
\newcounter{subtheorem}
\counterwithin{theoremgrp}{subtheorem}

\newcommand{\theoremgroup}{\refstepcounter{subtheorem}\refstepcounter{theorem}}

\DeclareMathOperator*{\argmin}{arg\,min}
\DeclareMathOperator{\sgn}{sgn}

\usepackage{float}
\usepackage{bbold}
\usepackage{balance}
\AtBeginDocument{%
  \providecommand\BibTeX{{%
    \normalfont B\kern-0.5em{\scshape i\kern-0.25em b}\kern-0.8em\TeX}}}

\copyrightyear{2022}
\acmYear{2022}
\setcopyright{acmlicensed}
\acmConference[AIES '22] {Proceedings of the 2022 AAAI/ACM Conference on AI, Ethics, and Society}{August 1--3, 2022}{Oxford, United Kingdom.}
\acmBooktitle{Proceedings of the 2022 AAAI/ACM Conference on AI, Ethics, and Society (AIES '22), August 1--3, 2022, Oxford, United Kingdom}
\acmPrice{15.00}
\acmDOI{10.1145/3514094.3534156}
\acmISBN{978-1-4503-9247-1/22/08}

\begin{document}
\fancyhead{}

\title{Explainability's Gain is Optimality's Loss? -- How Explanations Bias Decision-making}

\author{Charles Wan}
\affiliation{%
  \institution{Rotterdam School of Management, Erasmus University}
  \streetaddress{Burgemeester Oudlaan 50}
  \city{Rotterdam}
  \country{The Netherlands}
  \postcode{3062 PA}
}
\email{wan@rsm.nl}

\author{Rodrigo Belo}
\affiliation{%
  \institution{Nova School of Business and Economics, Universidade Nova de Lisboa}
  \streetaddress{Carcavelos Campus, Rua da Holanda, no. 1}
  \city{Carcavelos}
  \country{Portugal}
  \postcode{2775-405}
}
\affiliation{%
  \institution{Rotterdam School of Management, Erasmus University}
  \streetaddress{Burgemeester Oudlaan 50}
  \city{Rotterdam}
  \country{The Netherlands}
  \postcode{3062 PA}
}
\email{rbelo@rsm.nl}

\author{Leid Zejnilović}
\affiliation{%
  \institution{Nova School of Business and Economics, Universidade Nova de Lisboa}
  \streetaddress{Carcavelos Campus, Rua da Holanda, no. 1}
  \city{Carcavelos}
  \country{Portugal}
  \postcode{2775-405}
}
\email{leid.zejnilovic@novasbe.pt}

\renewcommand{\shortauthors}{Wan, Belo and Zejnilović}

\begin{abstract}
  Decisions in organizations are about evaluating alternatives and choosing the one that would best serve organizational goals.  To the extent that the evaluation of alternatives could be formulated as a predictive task with appropriate metrics, machine learning algorithms are increasingly being used to improve the efficiency of the process.  Explanations help to facilitate communication between the algorithm and the human decision-maker, making it easier for the latter to interpret and make decisions on the basis of predictions by the former.  Feature-based explanations' semantics of causal models, however, induce leakage from the decision-maker's prior beliefs.  Our findings from a field experiment demonstrate empirically how this leads to confirmation bias and disparate impact on the decision-maker's confidence in the predictions.  Such differences can lead to sub-optimal and biased decision outcomes.       
\end{abstract}



\maketitle

\section{Introduction} \label{introduction}
Decisions in organizations are about evaluating alternatives and choosing the one that would best serve organizational goals \citep{simon2013administrative} --- which of the two candidates to hire, whether to pursue this or that sales lead, how to allocate scarce resources among different intervention targets.  To the extent that the evaluation of alternatives could be formulated as a predictive task with appropriate metrics, machine learning algorithms are increasingly being used to improve the efficiency of the process.  Armed with big data and high performance computing, data scientists can train algorithms to learn complicated functions that map features to labels over some data distribution, where the labels form the evaluative criteria on which decisions are based. 

Because the algorithm learns a function in a high-dimensional data space, human interpretability has become an issue.  Both from the perspective of regulatory compliance \citep{european_commission_regulation_2016} and as a result of an organization's endogenous need to facilitate communication between the algorithm and human decision-makers \citep{simon2013administrative}, explanatory methods are now being implemented with increasing frequency.  Based on Shapley values from cooperative game theory, SHAP (SHapley Additive exPlanations) \citep{lundberg2020local} have become especially popular.  Similar to some of the other approaches, they have the semantics of causal models.  A subset of the features is given as the explanans for having contributed the most to generate the predicted outcome: because $X=x$, therefore $Y=y$.  While such semantics make an algorithm's predictions interpretable for the human decision-maker, they are not a faithful representation of the algorithm.  For example, the Shapley value of a feature will be smaller when other, highly correlated, features are included, rendering SHAP internally inconsistent with regard to highly correlated or causally dependent features \citep{kumar2020problems}.  

Human decision-makers form their own priors by imposing a causal structure on the data \citep{lu2008bayesian}.  A human causal model is likely to be sparse and biased, privileging strong and salient features \citep{lombrozo2007simplicity, lu2008bayesian}.  By explicating an algorithmic prediction in the semantics of causal models, explanatory methods such as SHAP could induce leakage from the decision-maker's prior beliefs.  More specifically, we hypothesize that the human decision-maker would feel more confident about a prediction when there is an overlap between her own causal model and the causal model implicit in the explanans.   

To test our hypothesis we designed and conducted a field experiment.  Counselors in a public employment agency had to view an algorithm's assessment of a candidate's risk for long-term unemployment, which they could either retain or override with their own assessment.  The counselors were then asked to rate their confidence in the final assessment.  The treatment group of counselors, in contrast to the control group, was shown SHAP for the algorithm's assessment.  Our results show that when strong causes in the counselor's own mental model are given as the explanans for the algorithm's prediction, the counselor's relative confidence in the final assessment increases. 

Our findings have important implications.  They provide empirical evidences of a mechanism by which explanations that serve to facilitate communication between the algorithm and human decision-makers could induce leakage from the decision-makers' prior beliefs.  This would lead to unwarranted confidence in the evaluations of some alternatives relative to others, making the decision-making process sub-optimal and biased.  In \nameref{discussion} we suggest possible solutions.

\section{Literature Review} \label{literature}

Machine learning is a general-purpose technology which organizations increasingly rely on to perform predictive tasks and boost productivity \citep{brynjolfsson2017can}.  The output it generates, however, is often opaque and difficult to interpret.  Machine learning as a black box creates "epistemic uncertainty and opacity" \citep{lebovitz2022engage}.  With the proliferation of machine learning models in all spheres of life there is a perceived need for mechanisms that can help users and other stakeholders to better understand algorithmic predictions.  The European Union's General Data Protection Regulation (GDPR) \citep{european_commission_regulation_2016}, for example, stipulates ``meaningful information about the logic involved'' in the case of automated decision-making.  Prior work shows that explanations can inculcate trust in automated decision aids \citep{dzindolet2003role} and expert systems \citep{ye1995impact}.  A number of methods for explaining machine learning predictions have been developed \citep{ribeiro2016should,lundberg2020local,montavon2018methods, russell2019efficient, galhotra2021explaining,molnar2020interpretable}. 

SHAP (SHapley Additive exPlanations) \citep{lundberg2020local}, in particular, have become very popular with a readily implementable Python package and exact algorithm for XGBoost.  The method converts features for a supervised learning problem into players in a cooperative game, with change in the predicted label as the payoff.  The Shapley value of an individual feature --- an individual player in the game --- is its marginal contribution to the label change, averaged over all possible subsets of features.  For each instance of prediction, SHAP rank the features by their Shapley values.  The most highly ranked ones are interpreted to exert the greatest influence on that particular instance of prediction.    

Recent theoretical work has shed light on the drawbacks of feature-based explanatory methods like SHAP \citep{camburu2020struggles, galhotra2021explaining, kumar2020problems}.  For our study two key insights due to \citep{kumar2020problems} are relevant.  Firstly, SHAP lacks internal consistency with respect to highly correlated or causally dependent features.  The inclusion of a highly correlated (in the extreme case, redundant) feature would change the Shapley values of the existing features and their ranking.  Secondly, the payoff --- the sum of the contributions of individual features to the label change --- is measured with reference to the unconditional expected value of the label: $f(X=x) - \mathbb{E}[f(X)]$.  The unconditional expected value of the label, $\mathbb{E}[f(X)]$, is an expectation over the data distribution --- it resides in the data space but is not necessarily a data instance that can be observed or imagined.  If we think of feature-based explanations as inducing the semantics of causal models --- features $X=x$ leading to label $Y=y$ --- both of these issues present problems for the semantic stability of SHAP.  The explanations are not invariant to the inclusion (or exclusion) of highly correlated or causally dependent features.  Their semantics also implicitly rely on the unconditional expected value of the label as a base case, which is hard for the human decision-maker to interpret or conceptualize \citep{kumar2020problems}.  A further problem with the semantics of causal models induced by feature-based explanations is their ontological and epistemological instability \citep{kohler2018eddie, kasirzadeh2021use}.  Social categories are liable to perspectival and interpretative variability.  Treating them as immutable properties hides causal assumptions that are not captured by the models' semantics \citep{kasirzadeh2021use}.    

Humans ``build models to capture key features of the world, to envisage possible scenarios and to generate explanations'' \citep{lagnado_2021}.  Causal reasoning is argued to be ``essential for machines to communicate with us'' \citep{pearl2018book}.  The semantics of causal models implicit in the explanans construe algorithmic output in a way that can be readily grasped by human decision-makers, even though this is far from a faithful translation of the actual mechanism.  By the same token, human decision-makers try to relate labels to features by imposing a causal structure on the data \citep{lagnado_2021}.  They form strong and sparse priors \citep{lu2008bayesian, lombrozo2007simplicity}.  When there is an overlap between the causal model induced by the explanation and the priors of the human decision-maker, confirmation bias can result \citep{GHASSEMI2021e745, nickerson1998confirmation}.  Our study is the first to empirically investigate this mechanism whereby the provision of explanations induces leakage from the human decision-maker's priors.  

Previous empirical work on human-algorithm interaction has focused on how humans react to algorithmic predictions or recommendations in laboratory settings \citep{binns2018s, dietvorst2015algorithm, dietvorst2018overcoming} and field experiments \citep{kawaguchi2020will, bundorf2019humans}.  A common setup is to apprise the human user of some piece of information about the algorithm --- e.g., its performance.  The psychological condition thereby induced is an exogenous source of variation.  Our study investigates how the human decision-maker's perception of an algorithmic prediction can be driven endogenously by the interaction between her prior beliefs and an explanation.

More broadly, our study is related to literature streams that explore the differences between predictions and decisions.  To predict is not the same as to make a decision.  In operations research, \citep{van2021data, loke2021decision} optimize decision-making or prescription rather than mere prediction.  In marketing, targeting customers with the highest churn probability is found not to be the optimal decision \citep{ascarza2018retention,lemmens2020managing}.  \citep{feiler2021noise} shows that overconfidence in product forecast can lead to bad decisions and profit loss.  In our study we demonstrate empirically that, whatever the accuracy and efficiency of a machine learning algorithm in evaluating alternatives with respect to some metric, explanations can introduce human biases and make the decision-maker more confident about some alternatives relative to others.  This relative increase in confidence due to the validation of a decision-maker's mental model is unwarranted and can lead to sub-optimal and biased decisions. 

Lastly, research on algorithmic bias \citep{buolamwini2018gender,mehrabi2021survey, barocas2017fairness} has mostly focused on data and data distributions that bias predictions.  Our work reveals a possibly novel mechanism by which bias can be introduced.  The provision of explanations induces leakage from the decision-maker's prior beliefs.  The disparate impact is not on the predictions themselves but on the confidence of the decision-maker in these predictions.  Such differences can lead to biased decision outcomes.

\section{Empirical Context} \label{context}

\subsection{Setup}

The empirical context for our field experiment is a public employment agency in a European country.  One of the agency's main tasks is to help unemployed individuals find a new job --- either by suggesting openings in their field or by enrolling them in training programs and re-qualifying them for new vocations.  Virtually all unemployed individuals register on this agency, since it is a requirement for getting unemployment benefits.  Counselors at said agency give unemployment risk assessments (low, medium or high) on unemployed individuals upon their registration --- with ``low'' (``high'') indicating a low (high) risk of long-term unemployment, defined as being involuntarily unemployed for twelve months or more.  We ran the field experiment from October 2019 to June 2020 and implemented a classification algorithm pre-trained on historical data using XGBoost \citep{chen2016xgboost}.  The pre-trained algorithm took as its input candidate features and returned as output a risk assessment.  Additionally we furnished the treatment group (``1'' in Table \ref{job_centers}) with SHAP (SHapley Additive exPlanation) \citep{lundberg2020local} for the predictions. 

The assignment of treatment was randomized at the level of job centers, with three in the control group and three in the treatment group.  Within a job center, counselors available at the time of registrations were allocated to the candidates.  It was not possible to randomize at the level of counselors due to technical limitations.  This is not a problem\footnote{This is relevant for the parallel trend assumption of diff-in-diffs.} because counselors at different job centers are comparable\footnote{See Appendix \ref{appendix c}.} and, furthermore, our empirical findings show that the observed effects are highly unlikely to be due to the heterogeneity of job centers\footnote{We discuss this in \nameref{results}.}.

During the field experiment, upon the registration of a candidate in a job center the counselor had to perform a risk analysis, run the model and view the algorithm's risk assessment\footnote{In addition to the new XGBoost classification algorithm counselors were also required to see a legacy model's assessment due to internal regulations.} --- which in the treatment group was supplemented with SHAP.  The SHAP indicated which six features had contributed the most to the algorithm's assessment.  For a ``high'' (``low'') risk prediction, the top six features that increased (decreased) long-term unemployment risk were shown; for a ``medium'' risk prediction, the top three features that, respectively, increased and decreased long-term unemployment risk were shown.  The counselor then had to make a final risk assessment, either retaining the algorithm's assessment or overriding it with her own.  Finally, she was asked to rate her confidence in the final risk assessment on a Likert scale of 1 to 5, with 5 being the most confident.  In addition to this mandatory procedure, the counselor could run the model at any time --- e.g., during consultation with a candidate.  

\subsection{Features and Data}

The XGBoost classification algorithm was trained on ninety-five features, of which eighty-nine pertained to the candidates and six to the job centers.  Candidates' features included demographic characteristics as well as those related to employment and past interventions.  

For each model prediction of ``high'' (``low'') risk, SHAP assigns a value of $+1$ ($-1$) to the six features that contribute the most to an increased (decreased) risk of long-term unemployment and ``NULL'' to the others.  For predictions of ``medium'' risk, SHAP assigns a value of $+1$ to the three features that contribute the most to an increased risk of long-term unemployment, $-1$ to the three features that contribute the most to a decreased risk of long-term unemployment and ``NULL'' to the others.  Our data also include counselors' final risk assessments and their confidence in them.  For a more detailed description of the experiment, see \citep{zejnilovic2021machine}.

\section{Theory and Hypotheses} \label{theory}

\subsection{The Human Mental Model}

To evaluate the effects of the interaction between explanations of algorithmic predictions and the human decision-maker's mental model, it is necessary to first identify the latter.  We consider two possible models of human judgment\footnote{They should be thought of as models of the human mental model.}.  They can be construed as behavioral models that map some informational input to a set of discrete choices. 

For both models the informational input consists of candidate features.  In the first model ($\mathcal{M}_{1}$), the final risk assessment is the discrete choice that the counselor has to make.  We observe only the input and the choice --- and not the actual cognitive process that connects the two.  Because humans interpret the world causally, impose a causal structure on data or events and predict outcomes by constructing internal causal models \citep{lagnado_2021}, we can assume that this is what happens here as well.  The cognitive process is some form of internal causal modelling --- the counselor builds and then uses a mental model to infer risk outcomes from candidate features.  For example, the counselor might regard the attribute of receiving Social Integration Income --- or personal circumstances associated with it --- as constituting a strong cause for prolonged unemployment.  This imputed causal relationship would be revealed by the counselor's final risk assessment.  The attribute of receiving Social Integration Income, ceteris paribus, would induce the counselor to a make an assessment of higher risk.  We hypothesize that we can find a subset of features as strong causes.
\theoremgroup
\begin{theoremgrp} \label{hyp1}
    There is a subset of candidate features which are strong causes in the sense of strongly inducing the counselor to make a final assessment of high, medium or low risk. 
\end{theoremgrp}

Humans build sparse models and privilege strong and salient causes \citep{lombrozo2007simplicity, lu2008bayesian}.  Algorithms, on the other hand, learn statistical relationships in a high-dimensional space.  They relate a label or outcome variable of interest to a large number of features and their higher-order interactions.  We can therefore also test the qualitative hypothesis that individual human models should be more sparse than the algorithm.  
\begin{theoremgrp} \label{hyp2}
    Individual human mental models are more sparse than the algorithm.
\end{theoremgrp} 
If we assume that there is commonality between the individual human models --- that individual counselors tend to find the same features important --- then the aggregate human model should be more sparse than the algorithm as well.    
\begin{theoremgrp} \label{hyp3}
    The aggregate human mental model is more sparse than the algorithm.
\end{theoremgrp}

In the second model ($\mathcal{M}_{2}$), the choice set for the counselor consists of 1) to retain the algorithm's assessment, 2) to adjust it upwards and 3) to adjust it downwards.  Conditional on the algorithm's assessment, a subset of candidate features induces or biases the counselor to consistently override the algorithm's assessment.  For example, an attribute of college education might consistently bias the counselor to adjust the risk assessment downwards, given an initial algorithmic assessment of high or medium risk.  If we can find such a subset of candidate features, they are likely to be strong causes in the counselor's mental model.
\begin{theorem} \label{hyp4}
    There is a subset of candidate features which consistently induces the counselor to adjust predicted risk upward or downward relative to the algorithm's assessment.
\end{theorem}

\subsection{The Semantics of Explanations}

Feature-based explanations in machine learning have the semantics of causal models.  They show the features that contribute to a predicted outcome (because $X=x$, therefore $Y=y$) or make counterfactual statements (were it not for $X$ taking on such values, $Y$ would not be such).  This facilitates communication with the human decision-maker but creates two possible channels of distortion.  Firstly, in compressing a complex algorithm into the readily communicable semantics of causal models, feature-based explanations introduce their own epistemic uncertainties and inconsistencies.  Specifically for SHAP, the semantics of the explanations are not invariant with respect to the inclusion of highly correlated or causally dependent features \citep{kumar2020problems}.  As an example of how this can go wrong, imagine a number of features that are causally dependent on a protected attribute.  The inclusion of these causally dependent features could decrease the rank of the protected attribute, making it seem less important.  This leads to a distorted and blinkered view of what is the root cause driving the predicted outcome.  Furthermore, the explanatory causal effect is with reference to the expected value of the label --- e.g., the assessed risk is higher than the expected risk (expectation taken over the training set distribution) because the candidate is of the 56+ age group.  However, it is not obvious how the human decision-maker should interpret the expected value of the label --- it might not be a realistic or even realizable data instance --- or calibrate it against her own expectation \citep{kumar2020problems}.

Secondly, the explanandum is the algorithm but is easily mistaken as some process in the real world.  For example, an explanation might show that the attribute of being of the 56+ age group is contributing to the assessment of high risk.  This means that the ``56+'' feature is causing the algorithm to produce an assessment of high risk.  But the explanation, given its semantics, is likely to be misinterpreted as implying a real causal relationship between being 56+ and a higher risk of long-term unemployment, which is not necessarily warranted. 

In short, explanations in machine learning facilitate communication between the algorithm and the human decision-maker via the semantics of causal models.  But they introduce distortions and lead to possible misinterpretation on the part of the human decision-maker.

\subsection{Confirmation Bias and Confidence}

Confirmation bias is an extensively studied cognitive phenomenon.  It "connotes
the seeking or interpreting of evidence in ways that are partial to existing beliefs,
expectations, or a hypothesis in hand" \citep{nickerson1998confirmation}.  The two necessary elements are, internally, some hypothesis and, externally, some evidence that can be interpreted as lending support to the hypothesis.  In the predictive task, the hypothesis is the mental model the counselor uses to predict risk outcomes from candidate features.  Potential evidence for the hypothesis comes from the explanations.  More specifically, the explanations' semantics of causal relationships induce the counselor to draw parallels with her own causal model.  This increases the likelihood of finding confirmatory evidence.  We therefore hypothesize that when there is a larger overlap between the strong causes of the counselor's mental model and the features which according to SHAP are contributing the most to the algorithm's risk assessment, and furthermore when SHAP are shown to the counselor, her confidence in the final risk assessment would increase, relative to when there is a smaller or no overlap or when SHAP are not shown.

\theoremgroup
\begin{theoremgrp} \label{hyp5}
    When there is a larger overlap between the strong causes of the counselor's mental model, as established in Hypothesis \ref{hyp1}, and the features which according to SHAP are contributing the most to the algorithm's assessment, and furthermore when SHAP are shown to the counselor, her confidence in the final risk assessment increases, relative to when there is a smaller or no overlap or when SHAP are not shown.  
\end{theoremgrp}

\begin{theoremgrp} \label{hyp6}
    When there is a larger overlap between the strong causes of the counselor's mental model that consistently induce her to adjust predicted risk upward (downward) relative to the algorithm, as established in Hypothesis \ref{hyp2}, and the features which according to SHAP are contributing the most to the algorithm's assessment, and furthermore when SHAP are shown to the counselor, her confidence in the final risk assessment increases, conditional on the counselor adjusting the risk upward (downward), relative to when there is a smaller or no overlap or when SHAP are not shown.  
\end{theoremgrp}

As shown by \citep{feiler2021noise}, overconfidence in forecast can lead to sub-optimal decisions.  We argue a more basic point.  Disparate levels of confidence brought about by leakage from the decision-maker's prior beliefs --- her own mental model --- can lead to sub-optimal and biased decisions.  As an example of what can go wrong, imagine two job candidates who are predicted to have equally good performance.  The explanation for candidate \#1 attributes the predicted outcome to the feature ``PhD'' whereas the explanation for candidate \#2 attributes the it to the feature ``Coursera certficate.''  Furthermore, assume that the decision-maker's prior belief is that having a PhD contributes to good performance.  Finding confirmatory evidence for her own mental model in the explanation for candidate \#1, she is likely to feel more confident about the prediction for candidate \#1 than the one for candidate \#2 and therefore decide to hire the former.  Even though the predictions themselves might not be biased, this differential assignment of confidence (or epistemic uncertainty) can lead to decisions that are sub-optimal, unfair and biased.

\section{Identification Strategy} \label{identification}

\subsection{Identifying the Human Model}

We use discrete choice models to identify the human mental model.  $\mathcal{M}_{1}$ maps a candidate's features to the final risk assessment:
\begin{equation*}
    \tag{$\mathcal{M}_{1}$}
    r = f(X \theta + \varepsilon),
\end{equation*}
\begin{equation*}
    r \in \{\text{``low''},\text{``medium''},\text{``high''}\},
\end{equation*}
where $X$ is the vector of candidate features, $r$ the final risk assessment and $\theta$ the model parameters.  The model parameters can be estimated using multinomial logit or probit regression.    

Specifically, we use LASSO multinomial regression with ten-fold cross-validation to constrain nonzero model coefficients to a subset of cardinality $q$ most responsible for driving the variance in $r$ \citep{hastie2009elements}:
\begin{equation*}
    \hat{\theta} = \argmin_{\theta} \mathcal{L}(\theta;\mathbf{X},\mathbf{r}; \lambda),
\end{equation*}
where $\mathcal{L}$ is the loss function and $\lambda$ is the penalty.  These features are the strong causes of $\mathcal{M}_{1}$.  By choosing the appropriate value of $\lambda$, we can limit the number of strong causes to $q$ so that there is be meaningful variation in overlap with the six features shown in SHAP.   

We also compare the sparsity of (individual and aggregate) human judgment with that of a similar model for the algorithm.  We estimate the respective models by setting $\lambda$ to values that minimize the multinomial deviances:
\begin{equation*}
    \lvert \mathcal{M}_{ind} \rvert <  \lvert \mathcal{M}_{agg} \rvert < \lvert \mathcal{M}_{algo} \rvert,
\end{equation*}
where the cardinality of a model $\mathcal{M}$ is the number of its nonzero coefficients.  
 
Since the counselors make their risk predictions after viewing the algorithm's assessments, $\mathcal{M}_{1}$ is strictly speaking a model of ``algorithm + human'' judgment.  However, as the counselors are responsible for the final risk assessments, we assume that $\mathcal{M}_{1}$ closely approximates a model of purely human judgment.

$\mathcal{M}_{2}$ maps a candidate's features to the counselor's adjustment to the algorithm's risk prediction: 
\begin{equation*}
    \tag{$\mathcal{M}_{2}$}
    y = g(X \beta + \varepsilon),
\end{equation*}
\begin{equation*}
    y \in \{\text{``decrease''},\text{``retain''},\text{``increase''}\} = \{-1,0,1\},
\end{equation*}
where $y \equiv \sgn(r_{human}-r_{algo})$ is the directional adjustment to the algorithm's prediction by the counsellor.  The model coefficients can likewise be estimated using multinomial logit or probit regression.  

We similarly use LASSO multinomial regression with ten-fold cross-validation to constrain nonzero model parameters to a subset of cardinality $q$ most responsible for driving the variance in $y$ \citep{hastie2009elements}:
\begin{equation*}
    \hat{\beta} = \argmin_{\beta} \mathcal{L}(\beta;\mathbf{X},\mathbf{y}; \lambda),
\end{equation*}
where $\mathcal{L}$ is the loss function and $\lambda$ is the penalty.  These features are the strong causes of $\mathcal{M}_{2}$.  By choosing the appropriate value of $\lambda$, we can limit the number of strong causes to $q$ so that there is be meaningful variation in overlap with the six features shown in SHAP.  

We make a distinction between upward biasing strong causes that prompt counselors to increase the risk prediction (relative to the algorithm's assessment) and downward biasing strong causes that induce them to decrease the risk prediction (relative to the algorithm's assessment).

\subsection{Estimating the Effects on Confidence}

We use a diff-in-diffs approach to estimate the effects of the interaction between the human mental model and the explanation on the counselor's confidence in the final risk assessment.

The first difference is between treatment (being shown SHAP) and control (not being shown SHAP).  The second difference is between, on the one hand, observations with a large overlap between $\mathcal{M}_{1}$ ($\mathcal{M}_{2}$) strong causes and the six features shown in SHAP and, on the other hand, observations with a small or no overlap between $\mathcal{M}_{1}$ ($\mathcal{M}_{2}$) strong causes and the six features shown in SHAP:  
\begin{align*}
    \lvert S \cap E \rvert \geq k, \\
    \lvert S \cap E \rvert \leq l,
\end{align*}
where $S$ is the set of $\mathcal{M}_{1}$ ($\mathcal{M}_{2}$) strong causes, $E$ the set of six features given by SHAP and $k$ and $l$ natural numbers between 0 and 6 with $k \geq l$.  We are constrained to choose $k$ and $l$ so that the two sets would have some minimum number of observations.  

\begin{figure}[H]
\centering
    \includegraphics[width=1\linewidth, trim=2.5cm 21cm 10cm 2.5cm]{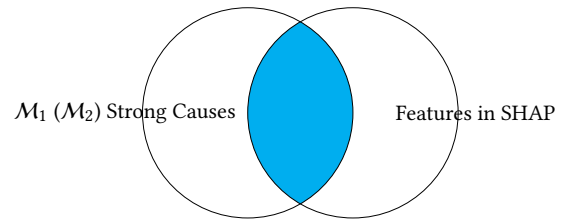}
    \caption{Cyan intersection is $\boldsymbol{S \cap E}$, where SHAP overlap with $\mathcal{M}_{1}$ ($\mathcal{M}_{2}$) strong causes.} 
    \label{fig:venn_diagram}
\end{figure}

We first test hypothesis \ref{hyp5}:
\begin{equation*}
    \Delta C_{treatment} - \Delta C_{control} > 0,
\end{equation*}
where
\begin{align*}
    \Delta C &= \frac{1}{n} \sum_{i}^{n} C_{i}*\mathbb{1}(\lvert S_{i} \cap E_{i} \rvert \geq k) \\ 
    &- \frac{1}{m} \sum_{j}^{m} C_{j}*\mathbb{1}(\lvert S_{j} \cap E_{j} \rvert \leq l), 
\end{align*}
$n$ and $m$ the numbers of observations in the respective sets.  $C$ is the counselor's confidence in the final risk assessment, $\Delta C_{treatment}$ and $\Delta C_{control}$ the respective differences in average confidence between the set of observations with a larger overlap and the set of observations with a smaller or no overlap.

We then test hypothesis \ref{hyp6}:
\begin{equation*}
    \Delta C_{treatment} - \Delta C_{control} > 0,
\end{equation*}
where
\begin{align*}
    \Delta C &= \frac{1}{n} \sum_{i}^{n} C_{i}*\mathbb{1}(\lvert S_{i} \cap E_{i} \rvert \geq k)*\mathbb{1}(y_{i} = \pm 1) \\
    &- \frac{1}{m} \sum_{j}^{m} C_{j}*\mathbb{1}(\lvert S_{j} \cap E_{j} \rvert \leq l)*\mathbb{1}(y_{j} = \pm 1), \\
\end{align*}
$n$ and $m$ the numbers of observations in the respective sets.  The effects are computed separately for when the counselors adjust predicted risk upward ($y$=+1) and for when they adjust downward ($y$=-1).

\subsection{Inference}

We use both regression and Fisher's randomization for statistical inference.  We run Fisher's randomization tests to construct p-values for the diff-in-diffs estimates.  Specifically, we make random draws from the observations and calculate $\Delta C_{treatment} - \Delta C_{control}$.  Statistical significance can then be tested by computing the $95$th and $99$th percentiles.

We also run the following regression model:

\begin{equation} \label{eq:1}
    C = \beta_{0} + \beta_{1} T + \beta_{2} O + \gamma(T*O) + \varepsilon,
\end{equation}
where $T=0$ ($T=1$) for the control (treatment) group; $O=0$ ($O=1$) for observations with at most $l$ overlapping features (at least $k$ overlapping features) between the strong causes of the human mental model and the six features shown in SHAP; and $C$ is the counselor's confidence in the final assessment.  We test the statistical significance of $\gamma$, which is the differential treatment effect $\Delta C_{treatment} - \Delta C_{control}$ estimated using diff-in-diffs. 

\section{Results} \label{results}

\subsection{Strong Causes in the Human Model}

We used the glmnet package in R \citep{friedman2010regularization} to run LASSO multinomial regressions.  By increasing the penalty $\lambda$ we constrain the set of features with nonzero coefficients in $\mathcal{M}_{1}$ to the seven listed in Table \ref{selected_m1}.  These are the candidate features that are the most responsible for driving variances in the final risk assessments.  Coefficients are for the three predicted classes.

\begin{table}[]
\caption{Strong causes in $\mathcal{M}_{1}$}
\label{selected_m1}
\begin{tabular}{rlccc}
  \toprule
 & features & low & medium & high \\ 
  \midrule
  1 & contract ended & 0.25 & 0.10 & -0.35 \\ 
  2 & social integration subsidy & -0.65 & -0.46 & 1.12 \\ 
  3 & age & -0.64 & 0.10 & 0.54 \\ 
  4 & previous subsidy suspensions & 0.08 & -0.02 & -0.06 \\ 
  5 & age group: 36-45 & -0.04 & 0.05 & -0.01 \\ 
  6 & age group: >56 & -0.03 & -0.23 & 0.26 \\ 
  7 & was LTU & -0.18 & 0.08 & 0.10 \\ 
  \bottomrule
\end{tabular}
\end{table}

Similarly, by increasing the penalty $\lambda$ we constrain the set of features with nonzero coefficients in $\mathcal{M}_{2}$ to the ones listed in Tables \ref{selected_m2_up} and \ref{selected_m2_dn}.  These are the candidate features that are the most responsible for making counselors deviate systematically from the algorithm's assessments.  Note that a binary feature can induce the counselor to either increase or decrease the risk assessment relative to the algorithm's risk prediction depending on the value it takes.  For example, $\texttt{ppe\_flag} = 1$ leads the counselor to adjust the predicted risk downward (relative to the algorithm) and $\texttt{ppe\_flag} = 0$ leads her to adjust it upward (relative to the algorithm).   

\begin{table}[]
\caption{Upward biasing strong causes in $\mathcal{M}_{2}$}
\label{selected_m2_up}
\begin{tabular}{rl}
  \toprule
 & upward biasing features \\ 
  \midrule
  1 & non-EU citizen \\ 
  2 & PALOP \\ 
  3 & education: 9th grade \\ 
  4 & no social integration subsidy \\ 
  5 & no personal employment plan\\ 
  6 & no previous LTU \\ 
   \bottomrule
\end{tabular}
\end{table}

\begin{table}[]
\caption{Downward biasing strong causes in $\mathcal{M}_{2}$}
\label{selected_m2_dn}
\begin{tabular}{rl}
  \toprule
 & downward biasing features \\ 
  \midrule
  1 & was student \\ 
  2 & education: college \\ 
  3 & desired job code: 2 \\
  4 & social integration subsidy \\ 
  5 & personal employment plan\\ 
  6 & first job  \\ 
  7 & was LTU \\ 
   \bottomrule
\end{tabular}
\end{table}

\subsection{Sparsity of the Human Model}

We ran LASSO multinomial regressions with $\lambda$'s minimizing multinomial deviances in order to compare the human model with the algorithm.  Table \ref{sparsity_models} shows that individual counselors do have sparse mental models and that even as an aggregate the human model is more sparse than a similarly estimated discrete choice model for the algorithm.  The cardinality of the discrete choice model for an individual counselor is considerably smaller than that of the discrete choice model for the algorithm.  The cardinality of the discrete choice model for the counselors as an aggregate is bigger, but still smaller than that of the discrete choice model for the algorithm. 

\begin{table}[]
\caption{The relative sparsity of the discrete choice models}
\label{sparsity_models}
\centering
\begin{tabular}{rc}
  \hline
 & cardinality \\ 
  \hline
$\lvert \mathcal{M}_{algo} \rvert$ & 145 \\ 
$\lvert \mathcal{M}_{agg} \rvert$ & 121 \\ 
$\lvert \mathcal{M}_{ind} \rvert$& 62\\
   \hline
\end{tabular}
\end{table}

\subsection{Differential Treatment Effects}

For $\mathcal{M}_{1}$ strong causes as listed in Table \ref{selected_m1}, we find the differential treatment effects $\Delta C_{treatment} - \Delta C_{control}$ to be statistically significant at the 0.01 or 0.05 level using Fisher's randomization.  The results are shown for three partitions of observations\footnote{\label{note1}The partition gives the second difference of the diff-in-diffs.}: 4 or more overlapping features vs. 0 overlapping feature; 4 or more overlapping features vs. 1 or less overlapping feature; 4 or more overlapping features vs. 2 or less overlapping 
features.   

\begin{table}[H]
\caption{Fisher's randomization results for $\mathcal{M}_{1}$}
\begin{tabular}{ccccc}
\toprule
k & l & $\Delta C_{t} - \Delta C_{c}$ & $95$th & $99$th  \\ \midrule
4 & 0 & 0.284*** & 0.066 & 0.093 \\ 
4 & 1 & 0.078*** & 0.035 & 0.049 \\ 
4 & 2 & 0.028** & 0.027 & 0.038 \\ 
\bottomrule
\addlinespace[1ex]
\multicolumn{3}{l}{\textsuperscript{***}$p<0.01$, 
  \textsuperscript{**}$p<0.05$}
\end{tabular}
\end{table}

For $\mathcal{M}_{2}$ upward biasing strong causes as listed in Table \ref{selected_m2_up}, we find the differential treatment effects to be statistically significant at the 0.01 level using Fisher's randomization.  The results are shown for two partitions of observations\footnote{See footnote \ref{note1}.}: 2 or more overlapping features vs. 0 overlapping feature; 2 or more overlapping features vs. 1 or less overlapping feature.   

For $\mathcal{M}_{2}$ downward biasing strong causes as listed in Table \ref{selected_m2_dn}, we find the differential treatment effects to be statistically significant at the 0.01 or 0.05 level using Fisher's randomization.  The results are shown for two partitions of observations\footnote{See footnote \ref{note1}.}: 2 or more overlapping features vs. 0 overlapping feature; 2 or more overlapping features vs. 1 or less overlapping feature.      
\begin{table}[H]
\caption{Fisher's randomization results for $\mathcal{M}_{2}$}
\begin{tabular}{rccccc}
\toprule
adjustment direction & k & l & $\Delta C_{t} - \Delta C_{c}$ & $95$th & $99$th \\ 
\midrule
upward & 2 & 0 & 0.265*** & 0.072 & 0.107 \\ 
upward & 2 & 1 & 0.157*** & 0.047 & 0.065 \\ 
downward & 2 & 0 & 0.099** & 0.077 & 0.111 \\ 
downward & 2 & 1 & 0.109*** & 0.069 & 0.095 \\ 
\bottomrule
\addlinespace[1ex]
\multicolumn{3}{l}{\textsuperscript{***}$p<0.01$, 
  \textsuperscript{**}$p<0.05$}
\end{tabular}
\end{table}

Regression model (\ref{eq:1}) as well as regression with overlap as a continuous variable give similar results\footnote{See Appendix \ref{appendix d}.}.  We observe that for both $\mathcal{M}_{1}$ strong causes and $\mathcal{M}_{2}$ upward biasing strong causes, a greater difference in overlap corresponds to a larger differential treatment effect.  This also indicates that the the differential treatment effect is highly unlikely to be due to the heterogeneity of job centers\footnote{Were the differential treatment effect solely due to intrinsic differences between the treatment and control groups of job centers, they would have to be different in such a way as to produce a larger differential treatment effect when the difference in overlap increases.}, justifying the parallel trend assumption of diff-in-diffs.        

\section{Discussion} \label{discussion}

Our empirical findings demonstrate that when there is a larger overlap between the human mental model and the model induced by the explanation and furthermore when the explanation is shown to the counselor, her confidence in the final assessment increases, relative to when there is a smaller or no overlap or when the explanation is not shown.  We also observe that the coefficients for treatment and overlap are negative when they are statistically significant\footnote{See regression results in Appendix \ref{appendix d}.}.  We interpret this as follows.  When there is a large overlap between features that are possibly important for the risk prediction and the counselor's prior beliefs but no explanation is given, the counselor's epistemic uncertainty increases and her confidence in the final assessment decreases.  On the other hand, when the counselor is shown the explanation but there is only a small or no overlap between the explanation and her prior beliefs, the counselor's epistemic uncertainty likewise increases and her confidence in the final assessment decreases.  Only when there is a large overlap and the counselor finds confirmatory evidence in the explanation given does the relative epistemic uncertainty decrease and confidence in the final assessment increase again.       

Therefore, leakage from prior beliefs produces disparate impact on the human decision-maker's epistemic uncertainty, depending on whether the overlap between the prior beliefs and the given explanation is large or small.  Even if the predictions themselves are optimal and bias-free, this mechanism makes the decision-maker more confident about some evaluations relative to others, leading to decisions that are sub-optimal and biased.  In our empirical setting, imagine that the counselor has to allocate scarce resources to two candidates judged to be equally at high risk of long-term unemployment.  Greater confidence in the risk prediction for one candidate compared to the other might very well lead to unfair allocation of resources.

In high-risk decision-making contexts such bail hearings, concordance between the human model and the causal model induced by the semantics of the explanations could lead to the illusion of understanding and gives a relative boost to confidence where it may not be epistemically warranted.  The leakage of prior beliefs into the decision-making process can also make any post hoc analysis or auditing difficult.  This would actually decrease the transparency of the system and make it less amenable to improvements.       

We offer some possible solutions.  Firstly, communication between algorithm and human decision-maker should shift away from the (implicit) semantics of causal models.  Human decision-makers should instead be trained to develop a general sense of the operative principles of the algorithm and its shortcomings.  Even when explanations are offered, decision-makers should be explicitly discouraged from thinking of them as corresponding to some causal process in the real world.  It might also be helpful to systematically expose human decision-makers to features and realized outcomes to help shape their priors.  Lastly, auditing should not be limited to the algorithm itself but needs to be extended to subsequent decision-making.  Decision-maker’s confidence in (or uncertainty about) a prediction and its impact on subsequent decisions should be carefully monitored.

\section{Acknowledgments}

\indent This work was funded by Fundação para a Ciência e a  Tecnologia (UIDB/00124/2020, UIDP/00124/2020 and Social   Sciences DataLab - PINFRA/22209/2016), POR Lisboa and POR  Norte (Social Sciences DataLab, PINFRA/22209/2016).

\setcitestyle{numbers} 
\bibliographystyle{ACM-Reference-Format}

\balance

\bibliography{bibliography}

\appendix 

\newpage

\section{Empirical Context} \label{appendix a}

\begin{table}[H]
\caption{Treatment assignment to job centers}
\label{job_centers}
\begin{tabular}{@{}c|cc|cc|c@{}}
\toprule
& \multicolumn{2}{c|}{pre-pilot} & \multicolumn{2}{c|}{pilot} & \multicolumn{1}{c}{} \\ \midrule
job center & regs. & appts./mo. & regs. & appts/mo. & \multicolumn{1}{c}{treated} \\ \midrule
1 & 11958 & 213 & 13139 & 169 & 0 \\
2 & 9406 & 191 & 10263 & 160 & 1 \\
3 & 3396 & 99 & 3743 & 72 & 0 \\
4 & 5717 & 110 & 6022 & 88 & 1 \\
5 & 3889 & 78 & 4379 & 69 & 0 \\
6 & 7016 & 135 & 7336 & 100 & 1 \\ \bottomrule
\end{tabular}
\end{table}

\section{Schematic for Identifying the Human Model}
\label{appendix b}

\begin{figure}[H]
\centering
    \includegraphics[scale=0.3]{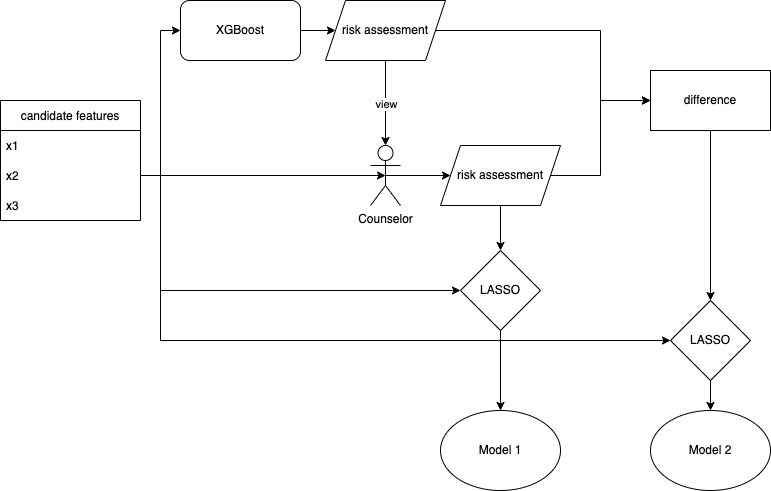}
    \caption{Two models of human judgment}
    \label{fig:schematic}
\end{figure}

\section{Comparison of Control and Treatment Groups}
\label{appendix c}

\begin{table}[H]
\caption{Wilcoxon test results}
\begin{tabular}{ccc}
\toprule
 & p-value & null hypothesis  \\ 
\midrule
age & 0.2861 & cannot reject \\ 
education & 0.1653 & cannot reject \\ 
tenure & 0.356 & cannot reject \\ 
gender & 0.4622 & cannot reject \\ 
candidates/mo. & 0.04025 & can reject \\
\bottomrule
\end{tabular}
\end{table}

\section{Results} \label{appendix d}

\subsection{Regression with overlap as a binary variable} 

\begin{table}[H]
\caption{Regression results for $\mathcal{M}_{1}$ strong causes with overlap as a binary variable}
\begin{tabular}{r c c c}
\toprule
\textbf{Variable} & \textbf{(k=4, l=0)} & \textbf{(k=4, l=1)} & \textbf{(k=4, l=2)}\\ 
\midrule
Intercept  &      3.878***   &  3.753***  & 3.731*** \\
        &      (0.026)     &      (0.011)   & (0.005) \\
Treatment  &      -0.250***   &      -0.044*** & 0.006 \\
        &      (0.037)     &      (0.015)   & (0.007)\\
Overlap &     -0.144***    &       -0.018   & 0.003\\
        &       (0.028)          &      (0.016) & (0.012) \\
Treatment*Overlap &  0.284***  &  0.078*** & 0.028*            \\
                 &  (0.040)   &  (0.021) &    (0.017)        \\
\bottomrule
\addlinespace[1ex]
\multicolumn{3}{l}{\textsuperscript{***}$p<0.01$, 
  \textsuperscript{**}$p<0.05$, 
  \textsuperscript{*}$p<0.1$}
\end{tabular}
\end{table}

\begin{table}[H]
\caption{Regression results for $\mathcal{M}_{2}$ upward biasing strong causes with overlap as a binary variable}
\centering
\begin{tabular}{r c c }
\toprule
\textbf{Variable} & \textbf{(k=2, l=0)} & \textbf{(k=2, l=1)} \\ 
\midrule
Intercept  &      3.813***   &  3.748***  \\
        &      (0.029)     &      (0.014)  \\
Treatment  &      -0.087***   &      0.023 \\
        &      (0.040)     &      (0.021)  \\
Overlap   &     -0.200***    &       -0.135***  \\
        &       (0.034)     &  (0.024) \\
Treatment*Overlap   &  0.268***  &  0.158*** \\
                 &  (0.049)   &  (0.035) \\
\bottomrule
\addlinespace[1ex]
\multicolumn{3}{l}{\textsuperscript{***}$p<0.01$, 
  \textsuperscript{**}$p<0.05$, 
  \textsuperscript{*}$p<0.1$}
\end{tabular}
\end{table}

\begin{table}[H]
\caption{Regression results for $\mathcal{M}_{2}$ downward biasing strong causes with overlap as a binary variable}
\centering
\begin{tabular}{r c c }
\toprule
\textbf{Variable} & \textbf{(k=2, l=0)} & \textbf{(k=2, l=1)} \\ 
\midrule
Intercept  &      3.789***   &  3.800***  \\
        &      (0.022)     &      (0.013)  \\
Treatment  &      -0.127***   &      -0.128*** \\
        &      (0.028)     &      (0.017)  \\
Overlap   &     -0.008    &       -0.019  \\
        &       (0.048)     &  (0.044) \\
Treatment*Overlap   &  0.103*  &  0.104* \\
                 &  (0.059)   &  (0.055) \\
\bottomrule
\addlinespace[1ex]
\multicolumn{3}{l}{\textsuperscript{***}$p<0.01$, 
  \textsuperscript{**}$p<0.05$, 
  \textsuperscript{*}$p<0.1$}
\end{tabular}
\end{table}

\subsection{Regression with overlap as a continuous variable}

\vspace{5mm}

\begin{table}[H]
\caption{Regression results for $\mathcal{M}_{1}$ strong causes with overlap as a continuous variable}
\centering
\begin{tabular}{r c}
\toprule
\textbf{Variable} \\ 
\midrule
Intercept  &      3.748*** (0.011)      \\
\addlinespace[1ex]
Treatment  &      -0.044*** (0.015)    \\
\addlinespace[1ex]
Overlap &     -0.007* (0.004)  \\
\addlinespace[1ex]
Treatment*Overlap &  0.025*** (0.006) \\
\bottomrule
\addlinespace[1ex]
\multicolumn{2}{l}{\textsuperscript{***}$p<0.01$, 
  \textsuperscript{**}$p<0.05$, 
  \textsuperscript{*}$p<0.1$}
\end{tabular}

\vspace{15mm}

\caption{Regression results for $\mathcal{M}_{2}$ upward biasing strong causes with overlap as a continuous variable}
\centering
\begin{tabular}{r c}
\toprule
\textbf{Variable} \\ 
\midrule
Intercept  &      3.806*** (0.021)      \\
\addlinespace[1ex]
Treatment  &      -0.056* (0.030)    \\
\addlinespace[1ex]
Overlap &     -0.084*** (0.014)  \\
\addlinespace[1ex]
Treatment*Overlap &  0.108*** (0.020) \\
\bottomrule
\addlinespace[1ex]
\multicolumn{2}{l}{\textsuperscript{***}$p<0.01$, 
  \textsuperscript{**}$p<0.05$, 
  \textsuperscript{*}$p<0.1$}
\end{tabular}

\vspace{15mm}

\caption{Regression results for $\mathcal{M}_{2}$ downward biasing strong causes with overlap as a continuous variable}
\centering
\begin{tabular}{r c}
\toprule
\textbf{Variable} \\ 
\midrule
Intercept  &      3.795***  (0.020)      \\
\addlinespace[1ex]
Treatment  &      -0.142*** (0.026)    \\
\addlinespace[1ex]
Overlap &     0.004  (0.021)    \\ 
\addlinespace[1ex]
Treatment*Overlap &  0.033 (0.026) \\
\bottomrule
\addlinespace[1ex]
\multicolumn{2}{l}{\textsuperscript{***}$p<0.01$, 
  \textsuperscript{**}$p<0.05$, 
  \textsuperscript{*}$p<0.1$}
\end{tabular}
\end{table}

\end{document}